\documentclass[prd
 ,twocolumn,a4paper
 ,secnumarabic%
,amssymb, amsmath,nobibnotes, aps,showpacs]{revtex4}
\expandafter\ifx\csname package@font\endcsname\relax\else
 \expandafter\expandafter
 \expandafter\usepackage
 \expandafter\expandafter
 \expandafter{\csname package@font\endcsname}%
\fi

\def\Journal#1#2#3#4{\emph{#1} {\bf #2}, #3 (#4)}

\newtheorem{theorem}{Theorem}[section]
\newenvironment{Theorem}{\begin{theorem}\begin{em}}{\end{em}
\end{theorem}}
\newtheorem{lemma}[theorem]{Lemma}
\newenvironment{Lemma}{\begin{lemma}\begin{em}}{\end{em}\end{lemma}}
\newtheorem{corollary}[theorem]{Corrolary}

\newtheorem{conjecture}[theorem]{Conjecture}
\newenvironment{Conjecture}{\begin{conjecture}\begin{em}}{\end{em}\end{conjecture}}
\newtheorem{definition}[theorem]{Definition}

\newtheorem{remark}[theorem]{Remark}

\newtheorem{example}[theorem]{Example}

\newtheorem{property}[theorem]{Property}

\newtheorem{proposition}[theorem]{Proposition}
\newenvironment{Proposition}{\begin{proposition}\begin{em}}{\end{em}\end{proposition}}

\def\beq{\begin{eqnarray}}
\def\eeq{\end{eqnarray}}
\def\best{\begin{eqnarray*}}
\def\eest{\end{eqnarray*}}
\def\bcom{}

\def\s5{\sqrt{5}}
\def\D{\mbox{D}}

\def\wph{\mu}

\def\di{\textrm{d}}
\def\tr{\textrm{tr}}

\def\iff{if and only if }

\def\p{\partial}
\def\tr{\textrm{tr}}

\def\a{\alpha}
\def\b{\beta}
\def\g{\gamma}
\def\G{\Gamma}
\def\d{\delta}
\def\e{\epsilon}

\def\t{\theta}

\def\l{\lambda}
\def\L{\Lambda}

\def\s{\sigma}

\def\w{\omega}
\def\W{\Omega}

\def\div{\mbox{div}}
\def\curl{\mbox{curl}}
\def\tr{\mbox{tr}}

\def\fpd#1#2{\frac{\partial #1}{\partial #2}}
\def\vf#1{\frac{\partial}{\partial{#1}}}

\newcommand{\vrk}{\hfill $\Box$}

\begin{document}

\title{Algebraically general, gravito-electric rotating dust}

\author{Lode Wylleman}%
\email{lwyllema@cage.ugent.be}
\affiliation{Faculty of Applied
Sciences TW16, Gent University, Galglaan 2, 9000 Gent, Belgium}
\date{\today}
\begin{abstract}
The class of gravito-electric, algebraically general, rotating
`silent' dust space-times is studied. The main invariant properties
are deduced. The number $t_0$ of functionally independent zero-order
Riemann invariants satisfies $1\leq t_0\leq 2$. The solutions may be
subdivided accordingly, and special attention is given to the
subclass $t_0=1$. Whereas there are no $\Lambda$-terms comprised in
the class, the limit for vanishing vorticity leads to two previously
derived irrotational dust families with $\Lambda>0$, and the
shear-free limit is the G\"{o}del universe.
\end{abstract}
\maketitle

\section{Introduction}

Dust space-times in general relativity, characterized as perfect
fluids with vanishing pressure $p=0$ and non-vanishing matter
density $\wph\neq 0$, with or without cosmological constant $\L$,
have attracted a lot of attention. Whereas irrotational dust models
form important
arenas for studying both the late universe
\cite{EllisNel,MaartensMatravers} and gravitational collapse
\cite{Bertschinger}, exact rotating dust solutions may serve to
describe phenomena on a galactic scale. Due to conservation of
momentum the dust flow lines must be geodesics, and the remaining
kinematic variables are the expansion scalar $\t\equiv u^a{}_{;a}$,
shear tensor $\s_{ab}\equiv u_{(a;b)}-\frac{1}{3}\t\,h_{ab}$
($\t_{ab}\equiv u_{(a;b)}$ being the expansion tensor) and vorticity
(or rotation) vector $\w^a\equiv \frac{1}{2}\e^{abc}u_{b;c}$, where
$u^a$ is the normalized dust 4-velocity, $h_{ab}\equiv
g_{ab}+u_au_b$ the projection tensor onto the comoving rest space,
$g_{ab}$ the space-time metric and $\e_{abc}\equiv\eta_{abcd}u^d$
the spatial projection of the space-time permutation tensor
$\eta_{abcd}$.

The Weyl tensor $C_{abcd}$, representing the locally free
gravitational field, is fully determined by its electric part
$E_{ab}$ and magnetic part $H_{ab}$ w.r.t.\ $u^a$, defined by
\begin{eqnarray}\label{EH}
    E_{ab} = C_{acbd}\,u^c\,u^d,\quad
    H_{ab}= \frac{1}{2}\e_{amn}C^{mn}{}_{bd}\,u^d.
\end{eqnarray}
When $H_{ab}=0, E_{ab}\neq 0$ the dust space-time is called
\emph{gravito-electric}, and the Petrov type is necessarily $I$ or
$D$. As the gravito-electric tensor is the general relativistic
generalization of the tidal tensor in Newtonian theory (see e.g.\
\cite{EllisDunsby}), such space-times were  alternatively termed
`Newtonian-like' in \cite{Maartens2}.

For Newtonian-like irrotational dust models, the covariant
propagation equations for $\s_{ab}$, $\t$, $E_{ab}$ and $\wph$,
(equations (\ref{PEdottheta}-\ref{PEdotsigma}) and
(\ref{PEdotE}-\ref{PEdotrho}) below, with $\w^a=0$) form an
autonomous first order system, leading to a set of \emph{ordinary}
differential equations when projected onto the common eigenframe of
$\s^a{}_{b}$ and $E^a{}_{b}$. No spatial gradients appear, such that
each fluid element evolves as a separate universe, once the
constraint equations are satisfied by the initial data. Such models
were therefore termed `silent' in \cite{Matarrese1}. The setting
looked very appealing towards numerical schemes and simulations in
astrophysical and cosmological context, e.g.\ for the description of
structure formation in the universe and the study of the
gravitational instability mechanism in general
relativity~\cite{Bruni,Croudace}, where a clear motivation for
taking $\omega^a=0,H_{ab}\approx 0$ was given in \cite{Matarrese2}.

However, in two independent papers~\cite{vanElst,Sopuerta}, the
propagation of the constraint $H_{ab}=0$ along $u^a$ was shown to
give rise to an  infinite chain of integrability conditions and
corresponding constraints on the above autonomous system. These
constraints are identically satisfied for Petrov type $D$, but led
the authors to conjecture that for the algebraically general case
only orthogonally spatially homogeneous solutions of Bianchi type I
(i.e., Saunders' cosmological models \cite{Saunders}) would be
allowed. In \cite{Vandenbergh1}, however, the conjecture was shown
to be false for strictly positive cosmological constant $\Lambda$,
by the explicit construction of two solution families characterized
by the presence of a geodesic space-like Weyl principal vector
field. The generalized conjecture that these two families exhaust
the inhomogeneous Newtonian-like ID models of Petrov type $I$ was
put forward in \cite{Wylleman}~\footnote{A recent attempt to prove
this conjecture~\cite{Apost} contains conceptual mistakes, as will
be commented on elsewhere.}.

The idea behind 1+3 covariantly silent models was extensively
explained and deepened in the introduction of \cite{vanElst}, while
in the discussion section of the same paper weaker conditions than
$H_{ab}=\w^a=0$ were indicated for establishing the silent property.
One of them is to allow for vorticity. This is a natural
generalization,
since silent perfect fluids must have a vanishing spatial gradient
of pressure~\cite{vanElst} and hence are non-rotating \emph{or} dust
(as follows from the Frobenius theorem \cite{Synge}).
However, it was questioned at the same time  whether this would
appreciably broaden the class of silent solutions, as severe
restrictions at first sight remain.

On the other hand, important classes of rotating dust models have
been found by assuming some kind of symmetry, or are algebraically
special. Respective examples are Winicour's
classification~\cite{Winicour} of stationary axisymmetric models
satisfying the circularity condition (see e.g.\ \cite{SKMHH}), and
the general rotating dust solution admitting time-like conformally
flat hypersurfaces with zero extrinsic and constant intrinsic
curvature as found by Stephani~\cite{Stephani} and generalized by
Barnes for non-zero $\Lambda$~\cite{Barnes}, which depends on seven
free functions of one coordinate and which turns out to be
gravito-electric and of Petrov type $D$. However, no algebraically
general and asymmetric rotating dust solutions have so far been
found~\footnote{To the best of the author's knowledge.}.

The above serves as a clear motivation to investigate the class of
Petrov type $I$, gravito-electric rotating dust models, which will
be denoted by ${\cal A}$, in more detail. In this paper the main
invariant properties of such models are deduced and a natural
subdivision based on the number $t_0$ of functionally independent
zero-order Riemann invariants is made. The paper is organized as
follows. Section 2 introduces the 1+3 covariant and Weyl principal
(WP) tetrad settings. The main result is contained in section 3,
where it is proved that each member of ${\cal A}$ necessarily has
$\Lambda\neq 0$ and possesses a space-like geodesic Weyl principal
vector field which is parallel to the vorticity, and where a
completed set of algebraic relations between the basic scalar
invariants (WP tetrad curvature components and connection
coefficients) is presented. In section 4 further invariant
properties are deduced and it is shown that $1\leq t_0\leq 2$. The
subclass corresponding to $t_0=1$ is partitioned into two families:
the first one comprises all non-expanding members of ${\cal A}$ (for
which necessarily $\L<0$), whereas the second one is characterized
by equal norms of the vorticity and shear tensors (for which
necessarily $\L>0$). Section 5 treats the zero vorticity, shear-free
and Einstein space limit cases. The final section summarizes the
results in a theorem and briefly discusses further features.

\section{Mathematical setting}

We use geometric units $8\pi G=c=1$ and the signature $(-,+,+,+)$
for space-time metrics. Round (square) brackets denote
(anti)symmetrization. By definition, a space-time belongs to ${\cal
A}$ if its metric $g_{ab}$ is a solution of the field equation
\begin{eqnarray}\label{fieldeq}
R_{ab}-\frac{1}{2}R\,g_{ab}+\Lambda\, g_{ab}=\wph\, u_a\,u_b,\quad
\wph\neq 0,
\end{eqnarray}
and if
\begin{eqnarray}\label{PErotdustdef}
H_{ab}=0,\quad T_2{}^3-T_3{}^2\neq 0,\quad   \w^a \neq 0,
\end{eqnarray}
where
\begin{eqnarray}\label{T2T3def}
T_{2}\equiv 6E^a{}_{b}E^{b}{}_{a},\quad T_3\equiv
36E^a{}_{b}E^{b}{}_{c}E^{c}{}_{a}
\end{eqnarray}
are convenient multiples of the quadratic and cubic Weyl invariants
$I$ and $J$~\cite{SKMHH}.\\

In the 1+3 covariant approach~\cite{Ehlers,Ellis}, the tensorial
quantities $\omega_a,\,\s_{ab},\,\t$, $E_{ab}$ and $\wph$ play the
role of the fundamental dynamical fields, while the parameter
$\Lambda$ and the covariantly constant fields $h_{ab}$ and
$\e_{abc}$ also enter the governing equations. A dot denotes
covariant derivation along $u^a$ (`time propagation'). For
convenience we define, for arbitrary tensors
$T^{ab\ldots}{}_{cd\ldots}$ and for any natural number $n$, the
operator ${\cal O}_n$ by
\begin{equation}\label{def:On}
{\cal O}_n(T)^{ab\ldots}{}_{cd\ldots}\equiv
\dot{T}^{ab\ldots}{}_{cd\ldots}+n\theta T^{ab\ldots}{}_{cd\ldots}.
\end{equation}
We further use the streamlined notation of \cite{Maartens1}. The
spatially projected, symmetric and trace-free part of a tensor
$S_{ab}$ is denoted by
\begin{equation}
S_{\langle ab \rangle}\equiv h_a{}^c h_b{}^d
S_{(cd)}-\frac{1}{3}S_{cd}h^{cd}h_{ab}=0.
\end{equation}
The covariant spatial derivative $D_a$, acting on arbitrary tensors
$T^{ab\ldots}{}_{cd\ldots}$, and the associated curl and divergence
(div) operators, acting on one-tensors $V_a$ and two-tensors
$S_{ab}$, are defined by
\begin{eqnarray}
\label{defD}& & D_e{T^{ab..}}_{cd..} = {h^a}_p{h^{b}}_{q}\cdots
{h_{c}}^{r}{h_{d}}^{s}\cdots {h_e}^f
{T^{pq..}}_{rs..;f},\\
    \label{divcurlV}& & \textrm{div}\, V = D_a V^a,\quad{\textrm{curl}\, V}_a=\e_{abc}D^b\,V^c,\\
  \label{divcurlS}& & {\textrm{div}\, S}_a = D^b S_{ab},\quad{\textrm{curl}\, S}_{ab}=\e_{cd(a}D^c {S_{b)}}^d.
\end{eqnarray}

The Ricci identity for $u^a$ and the second Bianchi identity, both
incorporating the field equation (\ref{fieldeq}) via the
substitution
\begin{equation}
R_{ab}=\left(\frac{\mu}{2}+\L\right)h_{ab}+\left(\frac{\mu}{2}-\L\right)u_a
u_b,
\end{equation}
are covariantly split into time propagation and constraint
equations. For members of ${\cal A}$ these are:
\begin{itemize}
    \item Time propagation equations:
\begin{eqnarray}
\label{PEdottheta} & &
{\cal O}_{\frac{1}{3}}(\t)=-\s_{ab}\s^{ab}+2\w^a\w_a-\frac{\wph}{2}+\L,\\
\label{PEdotsigma} & & {\cal O}_{\frac{2}{3}}(\s)_{ab}=-\s_{c\langle
a}\s_{b\rangle}{}^c
-\w_{\langle a}\w_{b\rangle}-E_{ab},\\
\label{PEdotw} & & {\cal O}_{\frac{2}{3}}(\w)_a=\s_{ab}\w^b,\\
\label{PEdotE} & & {\cal O}_1(E)_{ab}=3\s_{c\langle a}E_{b\rangle}{}^c-\w^c\e_{cd\langle a}E_{b\rangle}{}^d-\frac{\wph}{2}\s_{ab},\\
\label{PEdotrho} & & {\cal O}_1(\wph)=0.
\end{eqnarray}
\item Constraint equations:
\begin{eqnarray}
\label{PEdivs} & & \frac{2}{3}D_a\t-{\div\,\s}_a+{\curl\,\w}_a,\\
\label{PEdefH} & & {\curl\,\s}_{ab}+D_{\langle a}\w_{b\rangle}= H_{ab},\\
\label{PEdivw} & & \div\,\w = 0,\\
\label{PEdivE} & & \div\,E_a-\frac{1}{3}D_a\wph=0,\\
\label{PEdotH}& & \curl\,E_{ab}=0,\\
\label{PEdivH} & & {\cal C}^{(1)}{}_a\equiv
3E_{ab}\w^b-[\s,E]_a+\wph\w_a=0,
\end{eqnarray}
where $[\s,E]_a\equiv \e_{abc}\s^b{}_d E^{dc}$ is the one-tensor
dual to the commutator of $\s^a{}_b$ and $E^a{}_b$.
\end{itemize}
Notice that the set of first order equations
(\ref{PEdottheta}-\ref{PEdotrho}), augmented with
$\dot{\L}=\dot{h}_{ab}=\dot{\e}_{abc}=0$, indeed forms
a `silent' dynamical system, which is  due to $\dot{u}^a=\curl
H\,_{ab}=0$. Expressing $\dot{H}_{ab}=0$ and
${\textrm{div}\,H}_a=0$, on the other hand, respectively yields the
two constraints (\ref{PEdotH}) and (\ref{PEdivH}), the repeated time
propagation of which gives rise to two chains of integrability
conditions. In the next section attention will be focussed on the
second one, leading to a completed set of invariant relations for
${\cal A}$ under which
the first chain turns out to be identically satisfied.\\

For a generic model contained in ${\cal A}$, let ${\cal
B}\equiv(\p_0{}^a=u^a,\p_1{}^a,\p_2{}^a,\p_3{}^a)$ denote the  WP
tetrad, i.e., the essentially unique orthonormal eigenframe of
$E^a{}_b$ ($E_{12}=E_{13}=E_{23}=0$).
Below, capital Latin letters $A,B,\ldots$ denote WP tetrad indices
and run from 0 to 3, while Greek letters $\a,\b,\ldots$ are
$(\p_1{}^a,\p_2{}^a,\p_3{}^a)$-triad indices, run from 1 to 3 and
have to be read `modulo 3' (e.g.~$\s_{\a+1\a-1}=\s_{12}$ for
$\a=3$). Einstein's summation convention applies for both kinds of
indices. In particular we have $h_{\a\b}=\d_{\a\b}$, with $\d$ the
Kronecker delta symbol, and by convention we take
$\e_{123}=1$~\cite{Ellis}. The action of a vector field $X^a$ on a
scalar function $f$ is denoted by $X f$. The commutator coefficients
${\g^A}_{BC}$ and Ricci-rotation coefficients ${\Gamma^A}_{BC}$ of
${\cal B}$ are defined by
\begin{eqnarray}
    \label{commrel} [\p_B,\p_C]={\g^A}_{BC}\p_A,\quad
    (\p_B)^A{}_{;C}={\Gamma^A}_{BC}.
\end{eqnarray}
As for any rigid frame, the lowered coefficients
${\Gamma}_{ABC}=g_{AD}{\Gamma^D}_{BC}=\Gamma_{[AB]C}$ and
${\g}_{ABC}=g_{AD}{\g^D}_{BC}=\g_{A[BC]}$ are biunivocally related
by
\begin{eqnarray}
    {\gamma}_{ABC}=-2\Gamma_{A[BC]},\quad
    {\Gamma}_{ABC}=-\g_{[AB]C}+\frac{1}{2}\g_{CAB}.
\end{eqnarray}
Within the orthonormal tetrad formalism w.r.t.\ ${\cal B}$, the
matter density $\wph$ and suitable linear combinations of the
eigenvalues $E_{\a}$ of $E^a{}_b$ and of the rotation coefficients
$\G_{ABC}$ play the role of invariantly defined basic variables. In
the present paper we will use
\begin{eqnarray}
 \w_\a &=& \G_{0[\a-1\,\a+1]},\\
 \s_{\a+1\,\a-1} &=& \G_{0(\a-1\,\a+1)},\\
 \theta &=& {\Gamma^\b}_{0\b},\\
 h_\a &\equiv& \s_{\a+1\,\a+1}-\s_{\a-1\,\a-1}\nonumber\\&=& \Gamma_{\a+1\,0\,\a+1}-\Gamma_{\a-1\,0\,\a-1},\\
  x_\a&\equiv& E_{\a+1}-E_{\a-1},
\end{eqnarray}
where $h_1= -(h_2+h_3)$ and $x_1= -(x_2+x_3)$,
together with~\footnote{The $\Omega_\a$ are the non-zero components
of the angular velocity vector of the triad $(\p_\a)$ w.r.t.\ the
`inertial compass', see e.g.\ \cite{MacCallum} and references
therein.}
\begin{eqnarray}
   \Omega_\a  &\equiv& {\Gamma}_{\a-1\,\a+1\, 0},\\
    n_{\a}  &\equiv&   {\Gamma}_{\a+1\,\a-1\,\a},\\
    q_\a  &\equiv&  -{\Gamma}_{\a\,\a-1\,\a-1}= {\gamma}_{\a-1\,\a-1\,\a} \\
   r_\a  &\equiv& {\Gamma}_{\a\,\a+1\,\a+1}= {\gamma}_{\a+1\,\a\,\a+1}.
\end{eqnarray}
In terms of the $x_\a$ the quadratic and cubic Weyl invariants read
\begin{equation}
T_2=4(x_2^2+x_2x_3+x_3^2),\quad T_3=-4(x_1-x_2)(x_2-x_3)(x_3-x_1)
\end{equation}
and the Petrov type $I$ condition $T_2^3-T_3^2\neq 0$ becomes
$x_1x_2x_3\neq 0$.

The basic equations of the formalism are the commutator relations,
i.e., the first part of (\ref{commrel}) applied to scalar functions
$f$ (further denoted by com$_{BC}\,f$), and the projected Ricci and
second Bianchi equations. Within the ON formalism the Ricci
equations are further split into:
\begin{enumerate}
\item the first Bianchi equations $R_{A[BCD]}=0$, equivalent to the Jacobi identities
\begin{eqnarray}
    [\p_{[A},[\p_B,\p_{C]}]]^D =
    \partial_{[A}{\gamma^D}_{BC]}+{\gamma^F}_{[BC}{\gamma}^D_{A]F}=0.
\end{eqnarray}
Here $[\p_{[0},[\p_{\a+1},\p_{\a-1]}]]^0=0$ is the $\a$-component of
(\ref{PEdotw}), while $[\p_{[1},[\p_2,\p_{3]}]]^0=0$ is
(\ref{PEdivw});
\item the $\a\b$-components of (\ref{PEdotsigma}) and
(\ref{PEdefH});
\item the tetrad components of the field equation (\ref{fieldeq}).
Here the $00$-component is Raychaudhuri's equation
(\ref{PEdottheta}) and the $0\a$-component is (\ref{PEdivs}); the
$\a\b$-components
\begin{eqnarray}\label{einstein_triadequations}
&&\partial_C{\Gamma^C}_{(\a\b)}-\partial_{(\b}{\Gamma^C}_{\a) C}+
{\Gamma^C}_{DC}{\Gamma^D}_{(\a\b)}-{\Gamma^C}_{\a D}{\Gamma^D}_{\b C}\nonumber \\
&&= \left(\frac{\wph}{2}+\L\right)\delta_{\a\b}
\end{eqnarray}
are not covered by the Ricci-identity for $u^a$.
\end{enumerate}
The formulas
\begin{eqnarray}
\label{dotV}\dot{V}_\a =\, \partial_0 V_\a + \e_{\a\b\g}\Omega^\b
V^\g,\\ \dot{S}_{\a\b} =\, \partial_0 S_{\a\b} +
2\e_{\g\d(\a}\Omega^\g
{S_{\b)}}^\d,\\
D_\a V_\b =\, \partial_\a V_\b-V_\d\,\Gamma^\d{}_{\b\a},\\ D_\a
S_{\b\g} =\, \partial_\a S_{\b\g}-2S_{\d(\g}\Gamma^\d{}_{\b)\a}
\end{eqnarray}
and definitions (\ref{divcurlV}-\ref{divcurlS}) relate the covariant
differential operations to directional derivatives.

\section{Main result}

Starting from ${\cal C}^{(1)}{}_a=0$ one subsequently derives the
necessary conditions
\begin{eqnarray*}
{\cal C}^{(2)}{}_a&\equiv& {\cal O}_{\frac{5}{3}}({\cal
C}^{(1)})_a+\frac{1}{2}\s_{ab}{\cal
C}^{(1)\,b}+\frac{1}{6}\e_{abc}\w^b
{\cal C}^{(1)\,c}\nonumber\\
&\equiv&\frac{8}{3}\s_a{}^b E_b{}^c\w_c+\frac{10}{3}E_a{}^b
\s_b{}^c\w_c-\s_b{}^{c}E_c{}^{b}\w_a=0
\end{eqnarray*}
and
\begin{eqnarray*}
&&{\cal C}^{(3)}{}_a\,\equiv\, {\cal O}_{\frac{7}{3}}({\cal
C}^{(1)})_a\nonumber\\
&&\equiv \frac{4}{3}\s_a{}^b \s_b{}^c E_c{}^d
\w_d+\frac{35}{3}\s_a{}^b E_b{}^c \s_c{}^d \w_d+5E_a{}^b \s_b{}^c
\s_c{}^d \w_d\nonumber\\
&&-2\s_b{}^c E_c{}^d
s_d{}^b\w_a-7\s_c{}^{d}E_d{}^{c}\s_a{}^b\w_b+2\s_c{}^{d}\s_d{}^{c}E_a{}^b\w_b
\nonumber\\
&&-\frac{4}{3}\w^c\w_c E_a{}^b\w_b-\frac{5}{3}E_{bc}\w^b\w^c
\w_a-6E_a{}^b E_b{}^c\w_c\nonumber\\&&+E_b{}^c E_c{}^b\w_a -3\wph
\s_a{}^b \s_b{}^c\w_c+\frac{\wph}{2}\s_b{}^c
\s_c{}^b\w_a\nonumber\\&&-\frac{5}{3}\e_{abc}\w^b
E^c{}_d\s^d{}_e\w^e
+\frac{5}{3}E_a{}^b\e_{bcd}\w^c\s^d{}_e\w^e\nonumber\\&&-\frac{4}{3}\s_a{}^b\e_{bcd}\w^c
E^d{}_e\w^e+\w^b \e_{bcd} E^c{}_e\s^{ed}\w_a=0.
\end{eqnarray*}

The validity of these equations has been independently checked in an
\emph{unspecified} tetrad approach, hereby using the Maple computer
algebra package. Note that the expansion scalar $\t$ does not enter
the expressions, which is a great technical advantage in view of
later elimination processes.

In this section we will show that the chain of algebraic
integrability conditions generated by further time propagation of
${\cal C}^{(3)}{}_a=0$ \emph{terminates}.  We will eventually
describe its complete solution set both covariantly and in terms of
WP tetrad invariants.
This will be achieved by a number of propositions.

Denote
\begin{eqnarray}\label{eq:defQZ}
F_\a\equiv h_{\a+1}x_{\a-1}+h_{\a-1}x_{\a+1},\\
Z_\a\equiv
\wph(x_{\a+1}-x_{\a-1})+2(x_{\a+1}^2+x_{\a-1}^2).\label{eq:defZ}
\end{eqnarray}
Projection of ${\cal C}^{(1)}{}_a=0$ w.r.t.\ the WP tetrad gives
\begin{eqnarray}
\s_{\a+1\,\a-1}=-\frac{\wph-x_{\a+1}+x_{\a-1}}{x_\a}\w_a,\label{eq:divHproj}
\end{eqnarray}
and substituting this in the components of ${\cal C}^{(2)}{}_a=0$
yields
\begin{eqnarray}
\label{eq:S4}\w_1 x_{2}x_{3}F_1+2\w_2w_3x_1Z_1=0,\\
\label{eq:S5}\w_2 x_{3}x_{1}F_2+2\w_3w_1x_2Z_2=0,\\
\label{eq:S6}\w_3 x_{1}x_{2}F_3+2\w_1w_2x_3Z_3=0.
\end{eqnarray}

The following lemmas allow to draw quick conclusions in later
proofs. Lemma \ref{Lemma1} especially helps to avoid  explicit
calculations; for its proof we need (\ref{PEdotrho}) together with
the diagonal components of (\ref{PEdotE}), namely
\begin{eqnarray}
{\cal O}_1(x_2)=\left(x_2+x_3-\frac{\wph}{2}\right)h_2+x_2h_3,
\label{eq:O2x2}\\
{\cal O}_1(x_3)=-x_3h_2-\left(x_2+x_3+\frac{\wph}{2}\right)h_3.
\label{eq:O2x3}
\end{eqnarray}

\begin{Lemma}\label{Lemma1} Suppose $x_2,\,x_3$ and $\wph$ are
constrained by two relations\\ $F(x_2,x_3,\wph)=G(x_2,x_3,\wph)=0$,
where $F$ and $G$ are homogeneous polynomials with integer
coefficients and without common factors. Then either $h_2=h_3=0$ or
$\wph^2=T_2$.
\end{Lemma}

{\textsc Proof.} As $F$ and $G$ do not have a common factor, their
resultant w.r.t.\ $x_3$ is non-zero~\footnote{See e.g.\ \cite{CLO},
pp.\ 158.} and thus leads to at least one \emph{irreducible}
homogeneous polynomial relation  $P(x_2,\wph)=0$. Let $n$ be the
total degree of $P$. As $O_1(\wph)=0$, we obtain
$O_{n}(P(x_2,\wph))=\fpd{P}{x_2}(x_2,\wph){\cal O}_1(x_2)=0$. Now
$\fpd{P}{x_2}$ and $P$ are both homogeneous and cannot have a common
factor, since $\fpd{P}{x_2}$ has a strictly lower degree than $P$
and $P$ is irreducible. Hence $\fpd{P}{x_2}(x_2,\wph)=P(x_2,\wph)=0$
would lead to $(x_2,\wph)=(0,0)$, which is excluded. Thus ${\cal
O}_1(x_2)=0$, and an analogous reasoning based on the resultant of
$F$ and $G$ w.r.t.\ $x_2$ yields ${\cal O}_1(x_3)=0$. By
(\ref{eq:O2x2}-\ref{eq:O2x3}) this gives a system of two linear and
homogeneous equations in $(h_2,h_3)$.
Hence either $h_2=h_3=0$ or the determinant of the system matrix,
computed to be $(\wph^2-T_2)/4$, vanishes.\vrk

\begin{Lemma}\label{Lemma2} (a) If two $F_\b$'s vanish at the same
time, then $h_2=h_3=0$.\\ (b) Two $Z_\b$'s cannot vanish
at the same time.\\
(c) If, for fixed $\b$, $Z_\b=0$ then $h_2=h_3=0$.
\end{Lemma}

{\textsc Proof.} By cyclicity it is sufficient to prove (a) for
$F_2=F_3=0$, (b) for $Z_2=Z_3=0$ and  (c) for $Z_1=0$.\\
(a) $F_2=F_3=0$ forms a linear and homogeneous system in the
variables $(h_2,h_3)$, the determinant of which is constantly
proportional to $x_1^2$ and
hence cannot vanish. Thus $h_2=h_3=0$.\\
(b) Elimination of $\wph$ from $Z_2=Z_3=0$ yields
$x_1(4x_2^2+7x_2x_3+4x_3^2)=0$, contradictory to the Petrov type $I$
assumption.\\
(c) One first calculates that
\begin{eqnarray*}
{\cal
O}_2(Z_1)&=&(4x_2^2-4x_3^2+4x_2x_3+\wph(2x_3-x_2)-\wph^2/2)h_2\nonumber\\
&+&(4x_2^2-4x_3^2-4x_2x_3+\wph(2x_2-x_3)+\wph^2/2)h_3.
\end{eqnarray*}
Combining the assumptions $Z_1=0$ and $\w^a\neq 0$ with the
equations (\ref{eq:S4}-\ref{eq:S6}) it follows that
$F_1F_2F_3=0$. Now suppose that $(h_2,h_3)\neq (0,0)$. Then, for
fixed $k$, the determinant $D_k=D_k(x_2,x_3,\wph)$ of the linear and
homogeneous system ${\cal O}_2(Z_1)=F_k=0$ in the variables
$(h_2,h_3)$ should vanish. As the computed $D_k$ is not a multiple
of $Z_1$, lemma \ref{Lemma1} applies with $F=Z_1$ and $G=D_k$.
Because of the hypothesis only the possibility $\wph^2-T_2=0$
remains, but elimination of $\wph$ from this equation and $Z_1=0$
yields $x_1^2x_2x_3=0$, contradictory to the Petrov type $I$
assumption. Thus $h_2=h_3=0$.\vrk\\

The key step in the deduction is the following

\begin{Proposition}\label{Prop:w1w2w3=0} $\w_1\w_2\w_3=0$.
\end{Proposition}

{\textsc Proof.} Suppose on the contrary that $\w_1\w_2\w_3\neq 0$.
Firstly, if $h_2$ and $h_3$ both vanished, one would have that all
$F_\a=0$ and hence, by (\ref{eq:S4}-\ref{eq:S6}), all $Z_\a=0$,
which is impossible according to lemma \ref{Lemma2}(b). Thus
$(h_2,h_3)\neq (0,0)$, whence also $Z_1Z_2Z_3\neq 0$ by lemma
\ref{Lemma2}(c). Secondly, the equations (\ref{eq:S5}),
(\ref{eq:S6}) form a linear and homogeneous system in the variables
$(\w_2,\w_3)$, the determinant of which must vanish by the
hypo\-thesis. Making analogous observations for the couples
(\ref{eq:S4}), (\ref{eq:S5}) and (\ref{eq:S4}), (\ref{eq:S6}) one
arrives at
\begin{eqnarray}\label{eq:wasquare}
L_\a\equiv 4Z_{\a+1}Z_{\a-1}\w_{\a}^2-F_{\a+1}F_{\a-1}x_{\a}^2=0.
\end{eqnarray}
The key steps are now the following. Substituting
(\ref{eq:divHproj}) into the first component of ${\cal
C}^{(3)}{}_a=0$ one obtains an equation of the form
\begin{eqnarray}\label{eq:dotdotdivHproj}
(P_1\w_1^2+P_2\w_2^2+P_3\w_3^2+
Qx_1^2x_2^3x_3^3)\w_1-Rx_1\w_2\w_3=0,
\end{eqnarray}
where $P_1,\,P_2,\,P_3$ are homogeneous polynomials in
$(x_2,x_3,\wph)$ of total degree 7, while $Q$ and $R$ are
homogeneous in $(x_2,x_3,\wph)$ and quadratic, resp.\ linear
homogeneous in $(h_2,h_3)$. Now multiply (\ref{eq:dotdotdivHproj})
with $2Z_1$, add $RL_1$, divide the result by $\w_1$ and finally
eliminate the $\w_\a^2$'s by means of (\ref{eq:wasquare}).
Performing the analogous operations on the second and third
components of ${\cal C}^{(3)}{}_a=0$
one derives three equations $G_\a=0$, where
\begin{equation*}
G_\a \equiv
Q_{\a+1\a+1}h_2^2+Q_{\a+1\a-1}h_2h_3+Q_{\a-1\a-1}h_3^2+2Z_1Z_2Z_3,
\end{equation*}
$Q_{22},\,Q_{23}$ and $Q_{33}$ being homogeneous polynomials in
$(x_2,x_3,\wph)$ of total degree 5. Consistency of these equations
with $(h_2,h_3)\neq (0,0)$ and $Z_1Z_2Z_3\neq 0$ requires that
$F=0$, where
\begin{eqnarray*}
F\equiv -1107\wph^8+T_2\wph^6+85T_3\wph^5 -1510T_2^2\wph^4
-63T_2T_3\wph^3\nonumber\\+2(130T_2^3-3T_3^2)\wph^2
-14T_3T_2^2\wph-2T_2(4T_2^3-T_3^2),
\end{eqnarray*}
as is readily deduced by putting $h_2$ equal to 1 in $G_2-G_1$ and
$G_3-G_1$, and then computing the resultant w.r.t.\ $h_3$ of the
resulting polynomials, which yields $Z_1Z_2Z_3F=0$. By repeating the
same procedure for $G_2-G_1$ and ${\cal O}_8(F)$, which is linear
homogeneous in $(h_2,h_3)$ by (\ref{PEdotrho}) and
(\ref{eq:O2x2}-\ref{eq:O2x3}), one arrives at a relation
$G(x_2,x_3,\wph)=0$, where $G$ is a homogeneous polynomial of total
degree 18 which is not a multiple of $F$. With this $F$ and $G$ one
deduces from lemma \ref{Lemma1} that $\wph^2-T_2=0$. Calculating the
resultant of $\wph^2-T_2$ and $F$, however, one finds
$T_2^2(T_2^3-T_3^2)^2=0$ such that the Petrov type is O or D, which
yields the desired contradiction.\vrk

\begin{Proposition}\label{Prop1} The vorticity is parallel to a Weyl
principal vector, i.e., it is an eigenvector of $E^a{}_b$.
\end{Proposition}

{\textsc Proof.} By proposition \ref{Prop:w1w2w3=0}, we must show
that the case $\w_1=0,\,\w_2\w_3\neq 0$ is inconsistent. From this
hypothesis and (\ref{eq:S4}) we get $Z_1=0$. Hence, by
(\ref{eq:defZ}) with $\a=1$, $x_2\neq x_3$ and
\begin{equation}
\label{spec}\wph=2\frac{x_2^2+x_3^2}{x_3-x_2}.
\end{equation}
The $\a=1$-component of (\ref{eq:divHproj}) immediately yields
$\s_{23}=0$, while substitution of (\ref{spec}) into the
2- and 3-components gives
\begin{eqnarray}
\label{setA}\s_{12} = \w_3\frac{x_2+3x_3}{x_2-x_3},\quad \s_{13}
=\w_2\frac{3x_2+x_3}{x_2-x_3},
\end{eqnarray}
respectively. Likewise, the $23$-component of (\ref{PEdotE}) yields
$\W_1=0$, while substitution of (\ref{spec}) into the $31$- and
$12$-components gives
\begin{eqnarray}
\label{setW}\W_2 = 4\w_2 x_2\frac{x_2+x_3}{(x_2-x_3)^2},\quad \W_3
=4\w_3 x_3\frac{x_2+x_3}{(x_2-x_3)^2},
\end{eqnarray}
such that  $\W_2\W_3\neq 0$. Next, by (\ref{eq:S5}-\ref{eq:S6}) and
lemma \ref{Lemma2}(a), or by $Z_1=0$ and lemma \ref{Lemma2}(c), we
conclude that $h_2=h_3=0$. Propagating this along the
$\partial_0{}^a$ integral curves, on using the diagonal components
of (\ref{PEdotsigma}), and then substituting (\ref{setA}) and
(\ref{setW}) one gets
\begin{equation}
\label{setw2}\w_2^2
=\frac{(x_2+2x_3)(x_2-x_3)^3}{48x_2(x_2+x_3)^2},\quad \w_3^2
=\frac{(2x_2+x_3)(x_2-x_3)^3}{48x_3(x_2+x_3)^2},
\end{equation}
such that $(x_2+2x_3)(2x_2+x_3)\neq 0$. Applying ${\cal O}_1$ on the
left and right hand sides of these two equations one finds
$\t\w_2^2=\t\w_3^2=0$, whence $\t=0$. With the so far obtained
equations Raychaudhuri's equation (\ref{PEdottheta}) reduces to
\begin{equation}\label{spec2}
2x_2x_3+\Lambda(x_2-x_3)=0.
\end{equation}
Using (\ref{spec}), (\ref{setW}), (\ref{setw2}) and (\ref{spec2})
one derives the simple relations
\begin{eqnarray}
\label{eq:rho}&&\wph=2\Lambda+2(x_3-x_2),\\
\label{eq:E1} &&x_2-x_3=3(\W_2^2+\W_3^2+\Lambda),\\
&&x_2+x_3=3(\W_2^2-\W_3^2).
\end{eqnarray}
At this stage (\ref{PEdotsigma})-(\ref{PEdotrho}) reduces to the
vanishing of $\s_{13},\s_{12},\w_2,\w_3,x_2,x_3$ and $\wph$ under
$\p_0$, such that the above algebraic relations either propagate
consistently along the $\partial_0{}^a$ integral curves, in a
trivial way, or lead to $\p_0\W_2=\p_0\W_3=0$. However, propagating
them along the $\p_\a{}^a$ integral curves will lead to a
contradiction. Doing this for (\ref{eq:rho}) and mixing up with
equation (\ref{PEdivE}) and the off-diagonal components of
(\ref{PEdotH}) leads to
\begin{eqnarray}
\label{eq:p1x23}\p_1 x_2=2r_1x_3+q_1x_2,\quad \p_1 x_3=-r_1x_3-2q_1x_2,\\
\label{eq:p2x23}\p_2 x_2=-r_2x_1+q_2x_3,\quad \p_2 x_3=-r_2x_1-\frac{q_2x_3}{2},\\
\label{eq:p3x23}\p_3 x_2=q_3x_1+\frac{r_3x_2}{2},\quad \p_3
x_3=q_3x_1-r_3x_2.
\end{eqnarray}
The diagonal components of (\ref{PEdotH}) on the other hand reduce
to
\begin{equation}\label{T1T2}
(x_2+x_3)n_1+x_2n_2=0,\quad (x_2+x_3)n_1+x_3n_3=0.
\end{equation}
Two more algebraic relations follow by suitably combining the 2- and
3-components of (\ref{PEdivs}) with the 31- and 12-components of
(\ref{PEdefH}), respectively, and substituting (\ref{setA}), namely
\begin{eqnarray}
\label{T3}\w_3(x_2+x_3)q_1+\w_2(3x_2+x_3)n_2+2\w_2x_2n_3=0,\\
\label{T4}\w_2(x_2+x_3)r_1+2\w_3x_3n_2+\w_3(x_2+3x_3)n_3=0.
\end{eqnarray}
Propagation of (\ref{spec2}) along the $\p_1{}^a$ integral curves
leads to
\begin{equation}\label{T5}
x_2(2x_2^2+x_3^2)q_1+x_3(x_2^2+2x_3^2)r_1=0.
\end{equation}
The equations (\ref{T1T2})-(\ref{T5}) form a linear and homogeneous
system in $q_1$, $r_1$ and the $n_\a$.  Calculating the determinant
of the system matrix and substituting (\ref{setw2}) one finds
$(x_2-x_3)^3(x_2+2x_3)(2x_2+x_3)(x_2+x_3)(x_2^2+x_3^2)/16$, which
cannot vanish. Hence
\begin{equation}\label{q1r1ns=0}
q_1=r_1=n_1=n_2=n_3=0.
\end{equation}
Next, applying $\p_2$ and $\p_3$ to (\ref{spec2})  leads to
\begin{eqnarray}
\label{T6}x_3(x_2^2+2x_3^2)q_2+2(x_3-x_2)(x_2+x_3)^2r_2=0,\\
\label{T7}2(x_3-x_2)(x_2+x_3)^2q_3-x_2(2x_2^2+x_3^2)r_3=0,
\end{eqnarray}
respectively. The com${}_{01}(x_2-x_3)$ commutator relation becomes
\begin{equation}\label{T8}
\w_3x_3^3q_2-\w_2x_2^3r_3=0.
\end{equation}
The com${}_{01}(x_2+x_3)$ commutator relation is a combination of
(\ref{T6}), (\ref{T7}) and (\ref{T8}). However, by propagating the
first equation of (\ref{setw2}) along $\p_2{}^a$ and the second one
along $\p_3{}^a$ we  find
\begin{eqnarray*}
\label{e2w2}&&\p_2\w_2=\frac{(x_2-x_3)^2}{192(x_2+x_3)^3x_2^2\w_2}\times\\
&&\left(x_3(5x_2+x_3)(x_2+2x_3)^2q_2-4(x_2^2+4x_2x_3+x_3^2)r_2\right),\\
\label{e3w3}&&\p_3\w_3=\frac{(x_2-x_3)^2}{192(x_2+x_3)^3x_3^2\w_3}\times\\
&&\left(4(x_2^2+4x_2x_3+x_3^2)q_3+x_3(x_2+5x_3)(2x_2+x_3)^2r_3\right),
\end{eqnarray*}
respectively. Substituting this in (\ref{PEdivw}) one gets a new
independent relation which, together with (\ref{T6}-\ref{T8}), forms
a linear and homogeneous system in $q_2$, $r_2$, $q_3$ and $r_3$.
After substitution of (\ref{setw2}) the determinant of the system
matrix is a homogeneous polynomial in $(x_2,x_3)$. If it does not
vanish then $q_2=r_2=q_3=r_3=0$; if it does vanish then it follows
in combination with (\ref{spec2}) that $x_2$ and $x_3$ are constant.
Looking at (\ref{eq:p2x23}) and (\ref{eq:p3x23}) we conclude that in
any case
\begin{equation}\label{qr23=0}
q_2=r_2=q_3=r_3=0.
\end{equation}
Finally, substitution of (\ref{setA}-\ref{setw2}), (\ref{eq:rho}),
(\ref{q1r1ns=0}), (\ref{qr23=0}) and $\t=h_2=h_3$ into the
11-component of (\ref{einstein_triadequations}) yields
$x_2-x_3=3\Lambda$ and thus, by comparison with (\ref{eq:E1}),
$\W_2^2+\W_3^2=0$, which is in contradiction with $\W_2\W_3\neq
0$.\vrk \vspace{.3cm}

By proposition \ref{Prop1} the vorticity vector is parallel to e.g.\
$\p_1{}^a$, such that $\w_1\neq 0$ and
\begin{equation}\label{w23=0}
\w_2=\w_3=0.
\end{equation}
Then the 2- and 3-components of (\ref{eq:divHproj}) simply read
\begin{equation}\label{s1/23=0}
\s_{12}=\s_{13}=0
\end{equation}
and the 31- and 12-components of (\ref{PEdotE}) yield
\begin{equation}\label{W23=0}
\W_2=\W_3=0.
\end{equation}
With these specifications, the equations (\ref{eq:S5}-\ref{eq:S6})
are automatically satisfied, whereas (\ref{eq:S4}) gives $F_1=0$,
i.e.,
\begin{equation}\label{F1=0}
h_2x_3+h_3x_2=0.
\end{equation}
Now the 1-components of ${\cal C}^{(1)}{}_a$ and ${\cal
C}^{(3)}{}_a/(12\w_1)$ are of the form
\begin{eqnarray}\label{frul}
\wph\w_1+F_1, \quad \wph(h_2h_3+\s_{23}^2)+F_2,
\end{eqnarray}
where $F_1$ and $F_2$ do not contain $\wph$. Calculating one further
derivative ${\cal O}_{3}(C^{3})_a$ and dividing the 1-component by
$2\w_1$ one gets a condition of the form
\begin{equation*}
(h_3-h_2)[4(h_2h_3+\s_{23}^2)+9\w_1^2-12(h_2x_3+h_3x_2)]\wph+F_3=0,
\end{equation*}
with $F_4$ again independent of $\wph$, which thus can be obviously
eliminated by means of (\ref{F1=0}) and (\ref{frul}). But doing so
one miraculously arrives at
$(h_2x_3+h_3x_2)F_4+x_2x_3(h_3-h_2+\t)=0$, such that
\begin{equation}\label{theta11=0}
\t=h_2-h_3 \quad\textrm{i.e.}\quad \t_{11}=0.
\end{equation}
by the Petrov type $I$ assumption. Thus the restriction to $u^\perp$
of the expansion tensor $\t^a{_b}$ has a zero eigenvalue,
$\t_1=\s_1+\t/3=0$, as is seen from the representation matrix
\begin{equation}\label{repr tab}
{}[\t_{\a\b}]=\left[\begin{array}{ccc}0&0&0\\0&-h_3&\s_{23}\\0&\s_{23}&h_2\end{array}\right],
\end{equation}
w.r.t.\ the WP triad $(\p_1{}^a,\p_2{}^a,\p_3{}^a)$. Applying $\p_0$
to (\ref{theta11=0}) one gets
\begin{equation}\label{rho-x rel}
\wph=2(x_2-x_3+\L)\quad\textrm{i.e.}\quad \wph=-6E_{1}+2\Lambda.
\end{equation}
The $\p_2$- and $\p_3$-derivatives hereof, together with $x_2x_3\neq
0$, give
\begin{equation}\label{q2r3=0 PErotdust}
q_2=r_3=0.
\end{equation}
Together with (\ref{theta11=0}) this implies that \emph{the
space-times possess a geodesic space-like WP vector field}, in casu
$\p_1{}^a$. In combination with (\ref{rho-x rel}) the 23-components
of (\ref{PEdotE}) and (\ref{eq:divHproj}) yield
\begin{equation}\label{plopp}
\s_{23}=-\frac{x_2-x_3+2\Lambda}{x_1}\w_1,\quad
\Omega_1=-2\frac{(x_2+\Lambda)(x_3-\Lambda)}{x_1^2}\w_1.
\end{equation}
A further $\p_0$-derivative of (\ref{rho-x rel}) leads us back to
(\ref{F1=0}). Propagating (\ref{F1=0}) along $\p_0{}^a$ and
substituting (\ref{plopp}) gives
\begin{eqnarray}\label{ge}
8\Lambda\w_1^2(x_2+\L)(x_3-\L)-2x_2x_3(x_1^2-\Lambda \t^2)=0.
\end{eqnarray}
On using (\ref{F1=0}) and the first equation of (\ref{plopp}) this
may still be simplified to
\begin{equation}\label{gf}
(h_2h_3+\s_{23}^2-\w_1^2)\L+x_2x_3=0.
\end{equation}
As the Petrov type is assumed to be $I$, (\ref{ge}) or (\ref{gf})
implies that $\Lambda\neq 0$. The $\p_0$-derivative of (\ref{gf}) is
found to be identically satisfied under the already derived
equations.

Equivalently, one may start with the \emph{covariant} expression of
(\ref{theta11=0}), namely
\begin{equation}\label{Rot dust  basisbetr theta11=0}
\t_{ab}\,\w^b=0 \quad \textrm{i.e.}\quad
\s_{ab}\,\w^b=-\frac{\t}{3}\t\w_a,
\end{equation}
with $\w^a\neq 0$. Two covariant time derivatives lead to
\begin{eqnarray}
&&E_{ab}\w^b=-\frac{1}{6}(\wph-2\Lambda)\w_a,\label{Eomega}\\
&&\tr(\s E)=\frac{\t}{3}(\wph-2\Lambda),
\end{eqnarray}
which, given (\ref{Rot dust  basisbetr theta11=0}), is nothing but
(\ref{F1=0}) and (\ref{rho-x rel}). Now ${\cal C}^{(1)}{}_a=0$
simplifies to
\begin{eqnarray}\label{[s,E]}
[\s,E]_a=\frac{1}{2}(\wph+2\Lambda)\w_a,
\end{eqnarray}
the covariant time derivative of which may be reduced by the already
obtained equations to
\begin{eqnarray}\label{Rot dust trE2}
\tr\, E^2-\frac{1}{6}(\wph-2\L)^2-2\tr(\s^2 E)+\frac{2}{3}\t\tr(\s
E) \nonumber\\+\frac{1}{18}(\wph+4\L)(3\tr\,\s^2-2\t^2)
-2\L\w^a\w_a=0,
\end{eqnarray}
The time propagation of (\ref{Rot dust trE2}) onto the WP tetrad is
identically satisfied under the projections of (\ref{Rot dust
basisbetr theta11=0}-\ref{Rot dust trE2}).

We conclude that, for perfect fluids satisfying
(\ref{fieldeq}-\ref{T2T3def}), \emph{the complete set of covariant
equations generated by time propagation of (\ref{PEdivH}) is
equivalent to (\ref{Rot dust basisbetr theta11=0}-\ref{Rot dust
trE2}) and, after projection onto the Weyl principal tetrad, to
(\ref{w23=0}-\ref{F1=0}), (\ref{theta11=0}), (\ref{rho-x rel}),
(\ref{plopp}) and (\ref{gf})}.

\section{Further invariant properties and subdivision}

From here it is advantageous to use
\begin{equation}
U_1\equiv \G_{203}=\w_1+\s_{23},\quad V_1\equiv
-\G_{302}=\w_1-\s_{23}
\end{equation}
as auxiliary variables. This allows to rewrite (\ref{plopp}) as
\begin{equation}\label{x2x3+-Lambda UVW1}
(x_3-\Lambda)U_1=(x_2+\Lambda)V_1=x_1\W_1\equiv-[(x_2+\L)+(x_3-\L)]\W_1,
\end{equation}
which is a linear and homogeneous system in the variables
$(x_2+\L,x_3-\L)$. As we would get $x_1=-(x_2+x_3)=0$ if both these
variables were zero,  the determinant of the system should vanish:
\begin{equation}\label{U1V1W1}
U_1V_1+U_1\W_1+V_1\W_1=0,\quad\textrm{i.e.}\quad
\W_1=\frac{\s_{23}^2-\w_1^2}{2\w_1}.
\end{equation}
After some manipulation, the projections of the Bianchi identity and
Ricci identity for $u^a$ reduce to
\begin{eqnarray}
&&\p_0 x_2=-h_2(x_2+\L),\quad \p_0 x_3=h_3(x_3-\L),\label{nm1}\\
&&\p_0 h_2=-h_2^2-2\W_1 U_1-x_2,\label{nm2a}\\
&&\p_0 h_3=h_3^2+2\W_1 V_1-x_3,\label{nm2b}\\
&&\p_0 \w_1=-\t\w_1,\quad \p_0\s_{23}=-\W_1 h_1-\t\s_{23},\label{nm3}\\
&&\p_1 x_2=-q_1 x_2,\quad \p_1 x_3=r_1 x_3,\label{nm4}\\
&&\p_2 x_2=-r_2 x_1,\quad \p_3 x_3=q_3 x_1,\label{nm5}\\
&&\p_1 h_2=-q_1h_2-2n_1\s_{23}+n_3V_1,\label{nm6}\\
&&\p_1 h_3=r_1h_3-2n_1\s_{23}-n_2U_1,\label{nm7}\\
&&\p_1 U_1=-q_1 U_1+2n_1h_2,\quad \p_1 V_1=r_1 V_1-2n_1h_3,\label{nm8}\\
&&\p_2 h_2+\p_3 V_1=-r_2 h_1+2q_3\s_{23},\label{nm9}\\
&&\p_3 h_3+\p_2 U_1=q_3 h_1+2r_2\s_{23},\label{nm10}\\
&&q_1 V_1-r_1U_1+2n_1(h_2-h_3)=0,\label{nm11}\\
&&n_1x_1=n_2x_2=n_3x_3=0,\label{nm12}
\end{eqnarray}
while the 11-component and subtraction of the 22- and 33-components
of (\ref{einstein_triadequations}) yield
\begin{eqnarray}\label{p1q1}
&&\p_1 q_1=-q_1^2-2n_1n_3-(x_2+\L),\\
\label{p1r1}&&\p_1 r_1=r_1^2+2n_1n_2-(x_3-\L).
\end{eqnarray}
\vspace{.2cm}

 On calculating the covariant time derivative of the
remaining primary integrability condition (\ref{PEdotH}), one checks
that its projection w.r.t.\ the WP tetrad is identically satisfied
under the above equations. For this purpose one uses the `curl-dot'
1+3 covariant commutator relation applied to the two-tensor
$E_{ab}$, which in our situation becomes~\footnote{This commutator
relation was given in \cite{Maartens1}, formula (A18), for the
further subcase $\w^a=0$. In the general formula (B.14) of
\cite{vanElstPhD}, for imperfect fluids and with $H_{ab}$ and $u^a$
not necessarily zero, the terms corresponding to $+3H_{c\langle
a}S_{b\rangle}{}^c$ have the wrong sign.}
\begin{eqnarray*}
(\curl\,S)^\cdot_{ab}&=&\curl\,\dot{S}_{ab}-\frac{1}{3}\t\curl\,S_{ab}-\s_e{}^c\e_{cd(a}\D^e
S_{b)}{}^d\nonumber\\
&&-\D_{(a}S_{b)}{}^d\w_d+\w_{(a}\div\,S_{b)}.
\end{eqnarray*}

\vspace{.2cm} Further properties of ${\cal A}$ may be deduced.
Firstly, $[\s,E]_a=0\,(\Leftrightarrow \s_{23}=0)$ is impossible:
$\s_{23}=0\neq \w_1$, (\ref{nm3}) and (\ref{U1V1W1})  would give
$-h_1\equiv h_2+h_3=0$, and then (\ref{nm2a}-\ref{nm2b}) would yield
$x_1=0$ and hence Petrov type $D$. Secondly, the shear tensor, or
equivalently the expansion tensor, cannot be degenerate. As
$\s_{23}\neq 0$, this would imply $h_3h_3+\s_{23}^2=0$, which
expresses the vanishing of the determinant of the non-trivial 2 by 2
block in (\ref{repr tab}),
But taking the $\p_0$-derivative hereof and using (\ref{F1=0}),
(\ref{U1V1W1}) and $x_2x_3\neq 0$ would then yield
$h_2=h_3=\s_{23}=0$, a contradiction. Thirdly, (\ref{rho-x rel})
implies that for each member of ${\cal A}$ the number $t_0$ of
functionally independent zero-order Riemann invariants is at most 2.
The following proposition states this much more precisely.

\begin{Proposition} Within ${\cal A}$ one has $1\leq t_0\leq 2$,
with moreover $t_0=1$ \iff  one of the three mutually exclusive
possibilities $\t=0$, $U_1=0$ or $V_1=0$ is satisfied. The
cosmological constant is positive for
$U_1V_1\equiv\w_1^2-\s_{23}^2=0$ and negative for $\t=0$.
\end{Proposition}

\textsc{Proof.} We have $t_0\leq 1$ \iff the differentials $\di x_2$
and $\di x_3$ are algebraically dependent at each point. Then, in
particular,
\begin{equation}\label{p0p1x2x3}
\p_0 x_2\,\p_1 x_3-\p_1 x_2\,\p_0 x_3=0.
\end{equation}
From (\ref{F1=0}), (\ref{nm1}), (\ref{nm4}) and $x_2x_3\neq 0$ it
follows that this equation is identically satisfied in the case
where $\t=h_2-h_3=0$ (which implies $h_2=h_3=0$), and that it gives
\begin{equation}\label{inh}
(x_3-\Lambda)q_1+r_1(x_2+\Lambda)=0
\end{equation}
when $\t\neq 0$. In combination with (\ref{x2x3+-Lambda UVW1}) and
$x_1\neq 0$, (\ref{inh}) is equivalent to
\begin{equation}\label{q1V1r1U1}
q_1V_1+r_1U_1=0.
\end{equation}
Now equations (\ref{F1=0}), (\ref{gf}) and the first equation of
(\ref{x2x3+-Lambda UVW1}) may be solved for $x_2$, $x_3$ and
$\Lambda$, giving
\begin{equation}\label{x2x3LFG}
x_2=-\frac{FG}{2h_3\w_1},\quad x_3=\frac{FG}{2h_2\w_1},\quad
\L=\frac{FG^2}{4h_2h_3\w_1^2},
\end{equation}
where
\begin{equation*}
F\equiv h_2h_3-U_1V_1\neq 0,\quad G=h_2V_1+h_3U_1\neq 0.
\end{equation*}
From (\ref{F1=0}) and (\ref{nm11}-\ref{nm12}) one obtains
\begin{eqnarray}\label{n2n3q1r1}
n_2=\frac{q_1V_1-r_1U_1}{2h_2},\quad
n_3=-\frac{q_1V_1-r_1U_1}{2h_3},\\
n_1=\frac{q_1V_1-r_1U_1}{2(h_2-h_3)}.\label{n1q1r1}
\end{eqnarray}
Taking the $\p_1$-derivative of (\ref{q1V1r1U1}), using (\ref{nm8})
and (\ref{p1q1}-\ref{p1r1}), and substituting (\ref{x2x3LFG}) and
(\ref{n2n3q1r1}-\ref{n1q1r1}), one surprisingly finds
\begin{equation}
U_1V_1 FG\frac{4\w_1^2 r_1^2+\t^2V_1^2}{2h_2h_3\t\w_1^2}=0.
\end{equation}
Together with (\ref{x2x3LFG}) and $\t x_2x_3\neq 0$ it follows that
$U_1V_1=0$.

Conversely, if $\t=0$, i.e., $h_2=h_3=0$, then
(\ref{nm2a}-\ref{nm2b}) implies $x_2=-2\W_1U_1$ and $x_3=2\W_1V_1$,
such that in particular $U_1V_1\W_1\neq 0$. Inserting this into
(\ref{gf}) one finds that $\Lambda=-4\W_1^2<0$; whence, without loss
of generality,
\begin{equation}\label{U1V1W1 h2h3=0}
\lambda\equiv \sqrt{-\L},\quad \W_1=\frac{\l}{2},\quad
U_1=-\frac{x_2}{\l},\quad V_1=\frac{x_3}{\l}.
\end{equation}
Herewith, either of the equations (\ref{x2x3+-Lambda UVW1}) becomes
\begin{equation}\label{speceq h2h3=0}
2x_2x_3+(x_2-x_3)\l^2=0,
\end{equation}
which establishes that the zero-order Riemann invariants are
\emph{algebraically} dependent. On the other hand, if e.g.\ $V_1=0$
$(\s_{23}=\w_1)$ then $U_1\neq 0$, and we consecutively deduce
$x_3=\Lambda$ from (\ref{x2x3+-Lambda UVW1}), $x_2=-h_2h_3$ from
(\ref{gf}) and $\Lambda=h_3^2>0$ from (\ref{F1=0}). The case $U_1=0$
is equivalent to $V_1=0$ (switching of 2- and 3-axes) and leads to
$x_2=-\L, x_3=h_2h_3$ and $\Lambda=h_2^2>0$.

Finally, if $t_0$ was allowed to be zero, then $h_2=h_3=0$ by
(\ref{nm1}), (\ref{F1=0}) and $x_1x_2x_3\neq 0$. Taking the
combination $x_3\,\p_1q_1-x_2\p_1 r_1$, using
(\ref{p1q1}-\ref{p1r1}) and then (\ref{nm12}), and finally inserting
$q_1=r_1=0$ (as follows from (\ref{nm4})) we arrive at
$(n_1^2-\Lambda)x_1=0$, contradictory
to the fact that $\Lambda<0$ when $h_2=h_3=0$.\vrk\\

\vspace{.3cm} It follows from the proof that when $t_0=1$, the only
allowed functional relations ${\cal F}(x_2,x_3)=0$ are
$x_2=\pm\lambda^2$, $x_3=\pm\l^2$ or (\ref{speceq h2h3=0}), with
$\l$ a real constant. Thus, e.g., linear relations involving
\emph{both} $x_2$ and $x_3$ are inconsistent. In particular, all
members of ${\cal A}$ are of Petrov type $I(M^+)$ in the extended
Arianrhod-McIntosh Petrov classification~\cite{Arianrhod1}, i.e.,
the eigenvalues $E_{\a}$ are all non-zero.\\

For the non-expanding subclass of ${\cal A}$ one derives from
(\ref{rho-x rel}), (\ref{U1V1W1}) and (\ref{U1V1W1 h2h3=0}) that
\begin{eqnarray}
w=-8\W_1(\W_1+\w_1)=2(\w_1^2+\s_{23}^2)\left(1-\frac{\s_{23}^2}{\w_1^2}\right).
\end{eqnarray}
This is positive \iff $|\w_1|>|\W_1|$ and $\W_1\w_1<0$, or \iff
\begin{equation}\label{shearvort}
|\w_1|>|\s_{23}|\quad\Leftrightarrow\quad
\sqrt{\w_{ab}\w^{ab}}>\sqrt{\s_{ab}\s^{ab}},
\end{equation}
where $\w_{ab}=u_{[a;b]}=\e_{abc}\w^c$ is the vorticity tensor. Thus
\emph{the energy density is positive \iff the norm of the
vorticity tensor is larger than the norm of the shear tensor}.\\

On the other hand, the expanding $t_0=1$ family is characterized by
$|\w_1|=|\s_{23}|$.  When e.g.\ $V_1=0$ ($\s_{23}=\w_1$) one may
take $h_3=-\l$ without loss of generality, which gives $x_2=\l h_2$
and $x_3=\l^2>0$. Still note from (\ref{nm1}) that
$x_3=\textrm{const}$ is an equivalent characterization for this
situation. We further deduce $\W_1=0$ from (\ref{x2x3+-Lambda
UVW1}), $r_1=q_3=0$ from (\ref{nm4}-\ref{nm5}), $n_1=n_2=n_3=0$ from
(\ref{nm8}) and (\ref{nm12}), and finally $r_2=0$ by taking the
$\p_2$-derivative of $x_2=\l h_2$ and using (\ref{nm5}),
(\ref{nm9}). Herewith the remaining equations are identically
satisfied or determine consistent expressions for derivatives of the
remaining variables $h_2$, $\w_1$ and $q_1$.

The invariant integration for this family was performed in
\cite{Wylleman2}. The result is the line element
\begin{eqnarray}\label{lineelement}
\lambda^2 ds^2&=&-(dt-2y\,dz)^2+(dx+f_1(z)dz)^2\nonumber\\
&&+e^{2t}\left[dy+\left(f_2(z)+y^2+e^{-2t}\right)dz\right]^2\nonumber\\
&&+(\cos\,x-e^{-t})^2 f_3(z)^2 dz^2,
\end{eqnarray}
where $\lambda=\sqrt{\Lambda}$ plays the role of a constant scaling
factor and where $t$, $x$ and $y$ are invariantly defined
coordinates. The arbitrary scalar functions $f_1(z),\,f_2(z)$ and
$f_3(z)$ are invariantly defined; only when these are all constant
there is a continuous isometry group, which is one-dimensional and
generated by $\partial/\partial z$. Writing $R\equiv \cos x\,e^{t}$
one has
\begin{equation}
\w_1=\frac{\l}{f_3(z)R},\quad \wph=\frac{2\l^2}{R}, \quad \t =
\l\left(1+\frac{1}{R}\right),\label{wrho}
\end{equation}
such that the matter density and expansion scalar are positive for
$R>0$.

\section{Limit cases}

The special gravito-electric irrotational dust solutions found in
\cite{Vandenbergh1}, namely
\begin{eqnarray}\label{generalmetric}
\L
ds^2&=&-dt^2+dx^2+[e^{-t}+g_1(y)\textrm{cos}\,x]^2\,dy^2\nonumber\\&&+[e^t+g_2(z)\textrm{sin}\,x]^2 dz^2,\\
\L
ds^2&=&-dt^2+dx^2+[e^{-t}+e^{2t}dz^2\nonumber\\&&+g_1(y)\textrm{cos}(x+g_2(y))]^2\,dy^2,\label{specialmetric}
\end{eqnarray}
were precisely characterized by having a zero eigenvalue for the
expansion tensor or, equivalently, by possessing a geodesic
space-like Weyl principal vector field. In the above studied
rotating case these properties have been shown to be satisfied
automatically, the singled out WP vector field being moreover
parallel to the vorticity. Thus, surprisingly, \emph{${\cal A}$
consists exactly of the rotating generalizations of
(\ref{generalmetric}-\ref{specialmetric}), which are on their turn
contained as the irrotational limit solutions}. The metric
(\ref{lineelement}) is the rotating generalization of
(\ref{specialmetric}). As also follows from the analysis in
\cite{Vandenbergh1}, the relations (\ref{w23=0}-\ref{F1=0}),
(\ref{theta11=0}-\ref{q2r3=0 PErotdust}) and (\ref{plopp}-\ref{gf})
with $\w_1=0$ are valid for the metrics
(\ref{generalmetric}-\ref{specialmetric}) when taking
$\vf{t}\sim\p_0$ and $\vf{x}\sim\p_1$. Note from (\ref{ge}) that
$\L=\lambda^2>0$, where $\lambda$ again plays the role of an overall
scaling factor.\\

The shear-free limits $\s_{23}=h_2=h_3=0$ of ${\cal A}$ are
non-expanding and automatically satisfy (\ref{shearvort}). We take
(\ref{x2x3+-Lambda UVW1}) -- instead of (\ref{plopp}) -- as a
defining relation for ${\cal A}$, and derive from (\ref{nm2}),
(\ref{nm4}), (\ref{nm11}), (\ref{nm12}), (\ref{x2x3+-Lambda UVW1})
and (\ref{rho-x rel}), respectively, that $x_1=0$,
$n_2=n_3=q_1=r_1=0$, $x_2=\wph+\L=\w_1^2=-\L>0$ and
$2\W_1=-\w_1=\l$. This corresponds precisely to the G\"{o}del
solution~\cite{Godel}
\begin{equation}
-2\L \di s^2=\di x^2+\di y^2+\frac{1}{2}e^{2x}\di z^2-(\di t+e^x \di
z^2),
\end{equation}
which may indeed be interpreted as dust $p=0$ with cosmological
constant $\L$.

Nowhere in the reasoning leading to the results of the previous
section we have explicitly used that $\wph\neq 0$. Because of
(\ref{rho-x rel}) and the inconsistency of linear relations
involving both $x_2$ and $x_3$ (cf.\ supra) it follows that Petrov
type $I$ $\L$-terms cannot be contained in ${\cal A}$ as `limits'
$\mu=0$. Thus we have established:

\begin{Theorem} Purely electric, algebraically general Einstein spaces
for which the time-like Weyl principal vector field is geodesic and
rotating do not exist.
\end{Theorem}

Mars~\cite{Mars} characterized the Kasner
space-times~\cite{Kasner1,Kasner2} as the purely electric Petrov
type $I$ vacua ($\mu=\L=0$) with non-rotating and geodesic time-like
WP vector field. Combining  the above theorem with this result and
with further work on the non-rotating Einstein space case
$\mu=0\neq\L$~\footnote{In collaboration with N.\ Van den Bergh it
has meanwhile been shown  that there should be at least an Abelian
$G_2$ isometry group. This will be presented elsewhere.}, the
following conjecture may be stated:

\begin{Conjecture}\label{conj L-term}  The only purely electric, algebraically general
Einstein spaces with a congruence of \emph{freely falling} Weyl
principal observers are the Kasner models and their generalizations
including a cosmological constant.
\end{Conjecture}

\section{Conclusion and discussion}

The following theorem summarizes the results obtained in this paper.

\begin{Theorem}\label{Theorem: PE rotdust main property}
Consider the class ${\cal A}$ of algebraically general,
gravito-electric rotating dust space-times. For any member of ${\cal
A}$, the vorticity vector at each point is parallel to a Weyl
principal vector which is moreover geodesic, the shear tensor does
not commute with the Weyl electric tensor and is non-degenerate, the
Petrov type is $I(M^+)$ in the extended Arianrhod-McIntosh Petrov
classification, and the cosmological constant cannot vanish. The
curvature varibles $E_{ab}$, $\mu$, $\L$ and kinematic quantities
$\s_{ab}$, $\w^a$ and $\t$ are subject to the covariant algebraic
relations (\ref{Rot dust  basisbetr theta11=0}-\ref{Rot dust trE2}).
The number $t_0$ of zero-order Riemann invariants is either 1 or 2.
The subclass corresponding to $t_0=1$ splits into two separate
parts. The first part consists of all non-expanding solutions
($\t=0$); these are Petrov type $I$, shearing
ge\-ne\-ra\-li\-za\-tions of the G\"{o}del universe, with $\L<0$,
and the energy density is positive \iff the norm of the vorticity
tensor is strictly larger than the norm of the shear tensor.
The second part corresponds precisely to the solutions for which
these norms are equal, where now $\L>0$. The class does not allow
for Einstein space ($\L$-term) limits $\wph = 0$.
\end{Theorem}

Note that these statements have been derived for dust ($p=0$)
space-times with cosmological constant, but they remain valid,
\emph{mutatis mutandis}, for perfect fluids with constant pressure.
E.g., it has been proved that if an algebraically general,
gravito-electric, non-expanding but rotating perfect fluid model has
constant pressure $p$, then $p$ must be larger than the cosmological
constant present.

A second remark concerns  the eventual relationship between the
rotation of a congruence of observers $u^a$ in a general space-time
and the magnetic part of the Weyl tensor with respect to
it~\cite{Glass}. Speculations about such a connection stem from the
fact that the Bianchi `div \textbf{H}' equation (for perfect fluids
or $\L$-terms: equation (\ref{PEdivH}) \, with ${\textrm{div}\,H}_a$
instead of 0 in the right hand side), contains the `angular momentum
density' source term $(\wph+p)\w_a$ -- in contrast to the analogous
Maxwell equation~\cite{MaBa,Ellis2}. The link was verified and
affirmed, in some sense, for e.g.\ the van Stockum
solution~\cite{Bonnor}~\footnote{The rotating dust interior of the
van Stockum cylinder, when staying close to the axis, is purely
electric w.r.t.\ a certain congruence of \emph{non-comoving}
observers.} and the Bondi space-time~\cite{Herrera}. However,
examples are known of rotating gravito-electric perfect
fluids~\cite{Collins2,Sklavenites}. For such space-times the third
term in (\ref{PEdivH}) is exactly balanced by the first two terms.
E.g., in the case of the Stephani-Barnes gravito-electric dust
space-times of Petrov type $D$ mentioned in the introduction, the
rotation lies in the $E^a{}_b$-eigenplane and one has
$3E_{ab}\w^b=-[\s,E]_a=-\wph\w_a/2$. For the metrics of class ${\cal
A}$ considered in this paper, the rotation is an eigenvector of
$E^a{}_b$ and there is an exact balancing based on (\ref{Rot dust
basisbetr theta11=0}), (\ref{Eomega}) and (\ref{[s,E]}). In this
respect, $\wph=-2\L$ again leads to the homogeneous G\"{o}del
solution, where now the rotation vector of the dust is an
$E^a{}_b$-eigenvector corresponding the non-degenerate eigenvalue
$2\L/3$ and spans the axis of local rotational symmetry.

Another consequence of the analysis is the following restatement of
the generalized silent universe conjecture (as put forward in
\cite{Wylleman}): \emph{an algebraically general gravito-electric
dust space-time is either an orthogonally spatially homogeneous
Bianchi type $I$ (OSH BI) Saunders model~\cite{Saunders}, or
possesses a geodesic space-like $E^a{}_b$-eigenvector field (and
then belongs to ${\cal A}$ or the non-rotating limit families
(\ref{generalmetric}-\ref{specialmetric}))}. Conjecture \ref{conj
L-term} states that only OSH BI models are possible for the
corresponding $\L$-term limit.


By this investigation, the question by van Elst \emph{et al.},
whether ${\cal A}$ constitutes a broad class of solutions (cf. the
introduction), may be answered affirmatively. We have seen that
${\cal A}$ may be partitioned into three subclasses, one constituted
by the non-expanding members, one by the expanding members with
$t_0=2$ and one by the expanding members with $t_0=1$. Equation
(\ref{lineelement}) gives the general line element corresponding to
the third subclass, which provides a first explicit example of
algebraically general rotating dust with an at most one-dimensional
isometry group, depending on three free functions of one coordinate.

On the other hand, the non-linearity of some of the class-defining
algebraic relations hinders a transparent consistency analysis for
the other two subclasses, at least in an orthonormal approach based
on an $E^a{}_b$-eigentetrad. However, as somewhat hidden in such an
approach, it turns out that there is a significant geometric duality
between the dust four-velocity $u^a$ and the normalized rotation
vector $v^a\equiv \w^a/\sqrt{\w_b\w^b}$ at each space-time point:
both are geodesic Weyl principal vectors, the vorticity vector of
the one is parallel to the other~\footnote{The vorticity of
$\p_1{}^a$ may vanish.}, and the one is an eigenvector of the shear
tensor of the other. These properties are most naturally expressed
within a 1+1+2 covariant formalism, the first `1' standing for $u^a$
and the second `1' for $v^a$, whereas `2' expresses that one leaves
a $SO(2,\mathbb{R})$ rotational freedom in the orthogonal complement
of these vectors at each point. There is good hope that, on using a
complexified version of such a formalism, one can elegantly tackle
the general consistency problem for ${\cal A}$, and thereby
substantiate that also the remaining subclasses $t_0=2$ and
$t_0=1,\,\t=0$ contain a large number of metrics.

\end{document}